\documentclass[
reprint,
superscriptaddress,
aps,
prb,
floatfix,
]{revtex4-2}
\usepackage{listings}
\usepackage{tikz}
\usepackage{times}
\usepackage{graphicx}
\usepackage{float}
\usepackage{latexsym,amsmath,amssymb,bm,euscript}
\usepackage{color}
\usepackage{subfigure}
\usepackage{epstopdf}
\usepackage[colorlinks=true,linkcolor=blue,citecolor=blue]{hyperref}
\usepackage{soul}
\usepackage[normalem]{ulem}
\usepackage{mathrsfs}
\usepackage{amssymb}
\usepackage{algpseudocode}
\usepackage{algorithm}
\usepackage{amsmath}
\usepackage{braket}
\usepackage{booktabs}
\usepackage{chngcntr}
\usepackage{lettrine}
\usepackage{CJKutf8}
\usepackage{bbding}
\usepackage{xspace}
\usepackage{textcomp}
\usepackage{textcase}
\usepackage{setspace}
\usepackage[T1]{fontenc}
\usepackage{float}
\usepackage{graphicx}
\usepackage{multirow}
\usepackage{mathtools}
\usepackage{threeparttable}
\usepackage[vlined, ruled, algo2e, linesnumbered]{algorithm2e}
\usepackage{changes}
\usepackage{makecell}

\usetikzlibrary{shapes}

\DeclareUnicodeCharacter{2212}{\ensuremath{-}}

\renewcommand\parallel{\mathrel{/\mskip-2.5mu/}}

\begin{document}

\preprint{APS/123-QED}

\title{C-type antiferromagnetic structure of topological semimetal CaMnSb$_2$}

\begin{CJK}{UTF8}{gbsn}
\author{Bo Li
	}
 \affiliation{School of Physics, Beihang University, Beijing, 100191, China}
\author{Xu-Tao Zeng
	}
 \affiliation{School of Physics, Beihang University, Beijing, 100191, China}
\author{Qianhui Xu 
	}
 \affiliation{School of Physics, Beihang University, Beijing, 100191, China}
\author{Fan Yang
	}
 \affiliation{School of Physics, Beihang University, Beijing, 100191, China}
\author{Junsen Xiang
	}
 \affiliation{Beijing National Laboratory for Condensed Matter Physics, Institute of Physics, Chinese Academy of Sciences, Beijing 100190, China}
\author{Hengyang Zhong
	}
 \affiliation{Center for Advanced Quantum Studies and Department of Physics, Beijing Normal University, Beijing 100875, China}
\author{Sihao Deng
	}
 \affiliation{Spallation Neutron Source Science Center, Dongguan 523803, China}
\author{Lunhua He
	}
 \affiliation{Spallation Neutron Source Science Center, Dongguan 523803, China}
 \affiliation{Beijing National Laboratory for Condensed Matter Physics, Institute of Physics, Chinese Academy of Sciences, Beijing 100190, China}
\affiliation{Songshan Lake Materials Laboratory, Dongguan 523808, China}
\author{Juping Xu
	}
 \affiliation{Spallation Neutron Source Science Center, Dongguan 523803, China}
\author{Wen Yin
	}
 \affiliation{Spallation Neutron Source Science Center, Dongguan 523803, China}
\author{Xingye Lu
	}
 \affiliation{Center for Advanced Quantum Studies and Department of Physics, Beijing Normal University, Beijing 100875, China}
\author{Huiying Liu
	}
 \email{liuhuiying@pku.edu.cn}
 \affiliation{School of Physics, Beihang University, Beijing, 100191, China}
\author{Xian-Lei Sheng
	}
 \email{xlsheng@buaa.edu.cn}
 \affiliation{School of Physics, Beihang University, Beijing, 100191, China}
\author{Wentao Jin
	}
 \email{wtjin@buaa.edu.cn}
 \affiliation{School of Physics, Beihang University, Beijing, 100191, China}

\date{\today}

\begin{abstract}
	
Determination of the magnetic structure and confirmation of the presence or absence of inversion ($\mathcal{P}$) and time reversal  ($\mathcal{T}$) symmetry is imperative for correctly understanding the topological magnetic materials. Here high-quality single crystals of the layered manganese pnictide CaMnSb$_2$ are synthesized using the self-flux method. De Haas-van Alphen oscillations indicate a nontrivial Berry phase of $\sim$ $\pi$ and a notably small cyclotron effective mass, supporting the Dirac semimetal nature of CaMnSb$_2$. Neutron diffraction measurements identify a C-type antiferromagnetic (AFM) structure below $T\rm_{N}$ = 303(1) K with the Mn moments aligned along the $a$ axis, which is well supported by the density functional theory (DFT) calculations. The corresponding magnetic space group is $Pn'm'a'$, preserving a $\mathcal{P}\times\mathcal{T}$ symmetry. Adopting the experimentally determined magnetic structure, band crossings near the Y point in momentum space and linear dispersions of the Sb $5p_{y,z}$ bands are revealed by the DFT calculations. Furthermore, our study predicts the possible existence of an intrinsic second-order nonlinear Hall effect in CaMnSb$_2$, offering a promising platform to study the impact of topological properties on nonlinear electrical transports in antiferromagnets.

\end{abstract}

\maketitle

\end{CJK}

\section{\label{sec:level1}INTRODUCTION}

Topological materials have attracted considerable attention in past decades due to their promising applications \cite{1, 2, 3, 4, 5, 6, 7, 8, 9, 10}. The investigations of these materials generally begin with the calculation of their electronic band structures, followed by experimental probes such as angle-resolved photoemission spectroscopy (ARPES) and quantum oscillation measurements. These methods play a crucial role in examining the actual band structure and offer valuable insights into potential nontrivial topological features of these materials. Dirac semimetals, a class of topological semimetals, are crystalline solids characterized by points in momentum space where doubly degenerate (enforced by the combined $\mathcal{P} \times \mathcal{T}$ symmetry) bands cross.  For magnetic Dirac semimetals, the absence of $\mathcal{T}$ symmetry leads to the disappearance of band degeneracy according to the Kramers theory, if $\mathcal{P}$ symmetry remains. In other words, the Dirac point exists only if $\mathcal{P}$ symmetry is broken but $\mathcal{P} \times \mathcal{T}$ symmetry is contained \cite{11}. Accurate calculation of the band structure and determination of symmetries require precise knowledges about the magnetic structure.  

The layered manganese pnictides \textit{A}Mn\textit{X}$_2$ (\textit{A} = Ca, Sr, Ba, or Yb; \textit{X} = Sb or Bi) containing similar Bi/Sb square lattices are predicted to be topological semimetals hosting massless Dirac fermions \cite{12, 13, 14, 15, 16, 17}. Extensive neutron diffraction studies and DFT calculations were carried out to determine the ground-state magnetic structure of the Mn$^{2+}$ ions and accordingly the symmetry. The Mn moments couple as a G-type AFM structure in SrMnBi$_2$, BaMnSb$_2$ and BaMnBi$_2$ \cite{12, 14, 18}, while form the C-type AFM order in CaMnBi$_2$ and YbMnSb$_2$ \cite{12, 16}. The topologically nontrivial predictions of these compounds are supported by DFT calculations and ARPES measurements \cite{18, 19, 20, 21, 22}. While their magnetic structures are not entirely the same, they both exhibit a symmetry operation \{$-1'|0$\} within their magnetic space groups, thereby preserving the $\mathcal{P} \times \mathcal{T}$ symmetry, which is a necessary condition for hosting Dirac fermions. 

Within the \textit{A}Mn\textit{X}$_2$ family, CaMnSb$_2$ crystallizes in an orthorhombic (space group: $Pnma$) structure. The observation of strongly anisotropic magnetoresistances suggests a quasi two-dimensional electronic structure, and quantum oscillation measurements further imply the presence of nearly massless Dirac fermions in this compound \cite{23}. In addition, ARPES measurements conclusively indicate the existence of a Dirac-like band structure in CaMnSb$_2$ \cite{24}. In contrast, both infrared spectroscopic measurement and DFT calculation tend to categorize CaMnSb$_2$ as a topologically trivial insulator, based on the assumption that the Mn moments exhibit a G-type AFM order \cite{25}. However, it is worth noting that the information about the true magnetic ground state of CaMnSb$_2$ is still experimentally lacking so far, which motivates us to perform a neutron diffraction study as the microscopic magnetic probe.

In this study, we have conducted a neutron powder diffraction experiment on CaMnSb$_2$, to directly determine the ground-state magnetic structure of the Mn moments. It is found that a long-range C-type AFM order develops below the N\'{e}el temperature of $T\rm_{N}$ = 303(1) K, with the Mn moments aligned along the $a$ axis, as well supported by DFT calculations. The magnetic space group $Pn'm'a'$ preserves a $\mathcal{P}\times\mathcal{T}$ symmetry, probably hosting an intrinsic second-order nonlinear Hall effect in CaMnSb$_2$. Additionally, DFT calculations adopting the C-type AFM structure clearly reveal band crossings near the Y point and linear dispersions of the Sb $5p_{y,z}$ bands, together with the characterizations from de Haas-van Alphen (dHvA) oscillations, supporting CaMnSb$_2$ as a topologically non-trivial magnetic Dirac semimetal.

\begin{table}	
	\caption{Crystal data and structural refinements for single-crystal CaMnSb$_2$.}
	\label{tab1}
	\begin{tabular}
		{
			>{\small}p{5cm}
			>{\small}c
		}
		\toprule [1pt]
		Formula mass (g/mol) & 331.74 \\
		Crystal system & orthorhombic \\
		Space group, $Z$ & Pnma (No.62), 4 \\
		$a$ (\AA) & 22.153(12) \\
		$b$ (\AA) & 4.343(2) \\
		$c$ (\AA) & 4.328(2) \\
		$V$ (\AA$^3$) & 416.4(4) \\
		$T$ (K) & 273(2) \\
		$\rho$ (cal) (g/cm$^3$) & 5.292 \\
		$\lambda$ (\AA) & 0.71073 \\
		F (000) & 577 \\
		$\theta$ ($\deg$) & 3.68-28.49 \\
		Reflections collected & 1836 \\
		Absorption coefficient (mm$^{-1}$)& 16.506 \\
		Final R indices & R$_1$ = 0.2145, wR$_2$ = 0.4930 \\
		R indices (all data) & R$_1$ = 0.2149, wR$_2$ = 0.4931 \\
		Goodness of fit & 1.231 \\
		\bottomrule [1pt]
	\end{tabular}
\end{table}

\begin{table*}	
	\caption{Wyckoff positions, coordinates, occupancies, and equivalent isotropic displacement parameters for CaMnSb$_2$}
	\label{tab2}
	\begin{tabular}
		{
			p{1cm}
			>{\centering\arraybackslash}p{3cm}
			>{\centering\arraybackslash}p{2cm}
			>{\centering\arraybackslash}p{2cm}
			>{\centering\arraybackslash}p{2cm}
			>{\centering\arraybackslash}p{2cm}
			>{\centering\arraybackslash}p{3cm}
		}
		\toprule [1pt]
		Atom & Wyckoff site & $x$ & $y$ & $z$ & Occupancy & $U\rm_{eq}$ \\
		\midrule [1pt]
		Ca & 4c & 0.6130(7) & 0.2500 & 0.731(4) & 1.000 & 0.021(3) \\
		Mn & 4c & 0.7499(6) & 0.7500 & 0.728(3) & 1.000 & 0.019(3) \\
		Sb(1) & 4c & 0.6721(2) & 0.7500 & 0.2196(9) & 1.000 &  0.0119(14)\\
		Sb(2) & 4c & 0.4998(3) & 0.2500 & 0.2465(13) & 0.9443 & 0.0213(17) \\
		\bottomrule [1pt]
	\end{tabular}
\end{table*}

\section{\label{sec:level1}Methods}

CaMnSb$_2$ single crystals were grown using the self-flux method similar to that report in Ref. \onlinecite{23}. In our case, the centrifugation of the melt was performed at a slightly higher temperature of 640 K, after which shiny millimeter-sized platelike crystals were obtained. The as-grown crystals were characterized through x-ray diffraction (XRD) by $\theta$-2$\theta$ scans at room temperature, using a Bruker D8 ADVANCE diffractometer with Cu K$\alpha$ radiation ($\lambda$ = 1.5406 \AA). Rocking-curve scans of representative Bragg reflections were carried out with high-resolution synchrotron x-ray ($\lambda$ = 1.54564 \AA) at the 1W1A beamline at the Beijing Synchrotron Radiation Facility, China. The single-crystal XRD data were collected at 273(2) K with the Mo K$\alpha$ radiation ($\lambda$ = 0.71073 \AA) using a Bruker D8A diffractometer. The data reduction was done with the Bruker SAINT software package, and the structural refinement was done with the Bruker SHELXTL software package. Magnetization measurements on single-crystal samples were performed on a 7 T Quantum Design MPMS and a 14 T PPMS, respectively.

Neutron powder diffraction experiments were conducted on the time-of-flight (TOF) diffractometer GPPD (General Purpose Powder Diffractometer) and the total scattering TOF diffractometer MPI (Multi-Physics Instrument), respectively, at China Spallation Neutron Source (CSNS), Dongguan, China \cite{26, 27}. High-quality single-crystal samples of CaMnSb$_2$ with a total mass of $\sim$ 1.1 g were thoroughly pulverized into fine powders and loaded into a vanadium container, and the neutron diffraction patterns were collected at selective temperature points in between 5 K and 350 K.  Rietveld Refinements of all diffraction patterns were carried out using the GSAS \MakeUppercase{\romannumeral2} program suite \cite{28}.

First-principles calculations were performed on the basis of DFT using the generalized gradient approximation (GGA) in the form proposed by Perdew $\mathit{et}$ $\mathit{al}$. \cite{29}, as implemented in the Vienna $ab$ $initio$ Simulation Package (VASP) \cite{30, 31}. The energy cutoff of the plane-wave was set to 500 eV. The energy convergence criterion in the self-consistent calculations was set to \textcolor{black}{10$^{-7}$} eV. A $\Gamma$-centered Monkhort-Pack $k$-point mesh with a resolution of 2$\pi$\texttimes{}0.04\textcolor{black}{{} \AA{}$^{-1}$} was used for the first Brillouin zone sampling. To account for the correlation effects for Mn, we adopted the GGA + $\mathit{U}$ method \cite{32} with the value of $\mathit{U}$ = 5 eV. 

\section{\label{sec:level1}RESULTS AND DISSCUTION}

\begin{figure}

	\centering
	\includegraphics[scale=0.7]{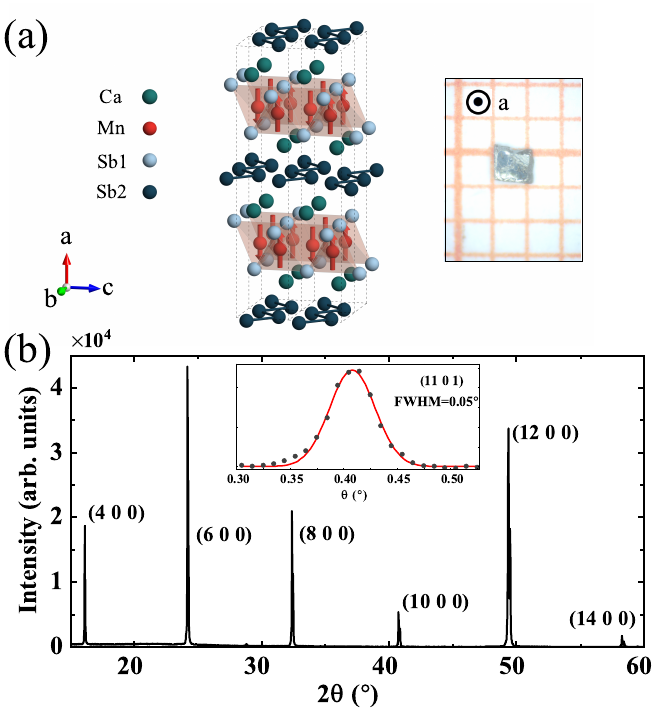}
	\caption{(a) The crystal structure of CaMnSb$_2$, with the experimentally determined C-type AFM spin configuration of Mn illustrated by red arrows, and a photo of the as-grown single crystal. (b) The XRD $\theta$-2$\theta$ scan of the single-crystal sample, with the inset representing a sharp rocking-curve scan. }
\end{figure}

Refinement of the single-crystal XRD data indicates an orthorhombic crystal structure for CaMnSb$_2$ (space group $Pnma$),  with the lattice parameters $a$ = 22.153(12) Å, $b$ = 4.343(2) Å, and $c$ = 4.328(2) Å at 273(2) K  (see Table \ref{tab1}), which agrees well with those reported in previous studies \cite{23, 33}. The refined atomic parameters are listed in Table \ref{tab2} and the corresponding crystal structure is illustrated in Figure 1(a). The atomic ratio between Ca, Mn and Sb is determined to be 1 : 1 : 1.94, very close to the stoichiometry of CaMnSb$_2$ with a slight deficiency of Sb(2) sites. 

The $\theta$-2$\theta$ scan of the CaMnSb$_2$ single crystal in Fig. 1(b) coincides well with the ($H$ 0 0) reflections, indicating that the normal direction of the platelike crystal is its crystallographic $a$ axis and the natural cleavage surface is the $bc$ plane. The narrow peak width (full width at half maximum, FWHM) of $\sim$ 0.05$^\circ$ for the rocking-curve scan of the (11 0 1) reflection shown in the inset of Fig. 1(b) suggests good quality of the
as-grown single-crystal samples.
	
\begin{figure}[h]
	\centering
	\includegraphics[scale=0.7]{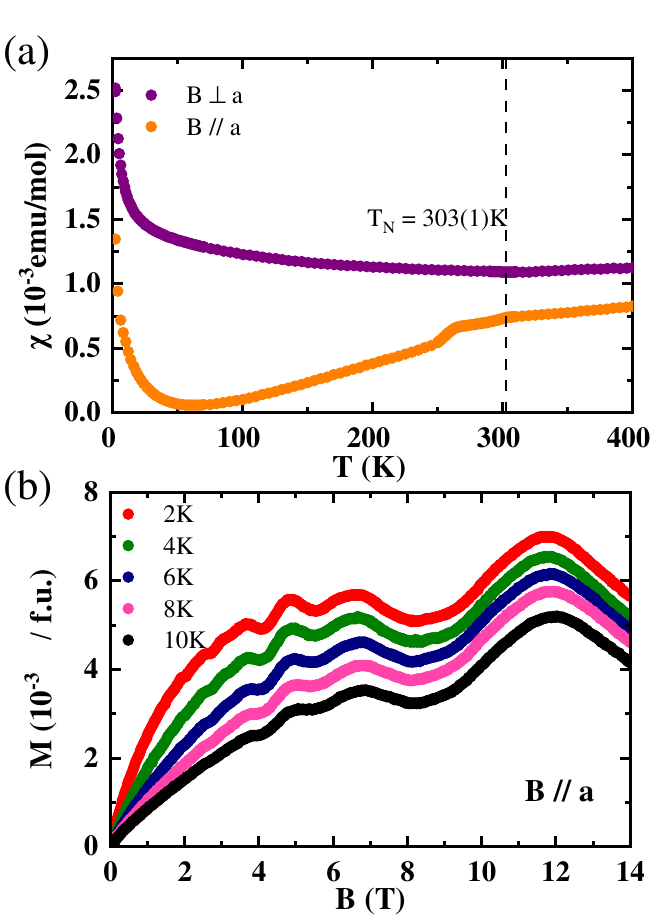}
	\caption{(a) The temperature dependence of dc magnetic susceptibility ($\chi$) of the CaMnSb$_2$ single crystal, measured in a magnetic field of 1 T applied along the $a$ axis ($B$$\parallel$$a$) or the $bc$ plane ($B$$\bot$$a$), respectively, in the zero-field-cooling (ZFC) mode. (b) The isothermal magnetization ($M$) curves measured up to 14 T  for $B$$\parallel$$a$ at different temperatures.}
	
\end{figure}

\begin{figure}
	\centering
	\includegraphics[scale=0.78]{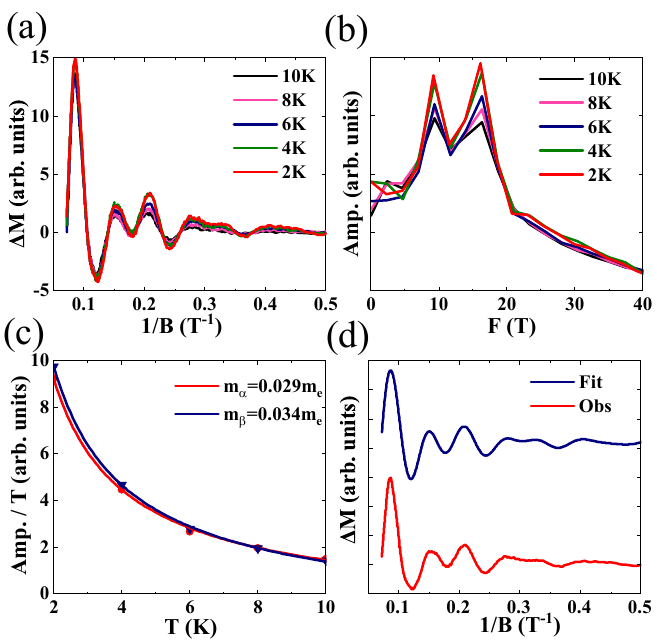}
	\caption{(a) Reciprocal magnetic field (1/$B$) dependence of the oscillatory parts $\Delta$$M$ at different temperatures. (b) FFT results of the dHvA oscillations. (c) Temperature dependence of the oscillating amplitude (Amp.) for two frequencies and the fittings. (d) Comparison between the experimental data of $\Delta$$M$ and the fitting using the Lifshitz-Kosevich formula.  }
\end{figure}

\begin{figure*}
	\centering
	\includegraphics[scale=0.65]{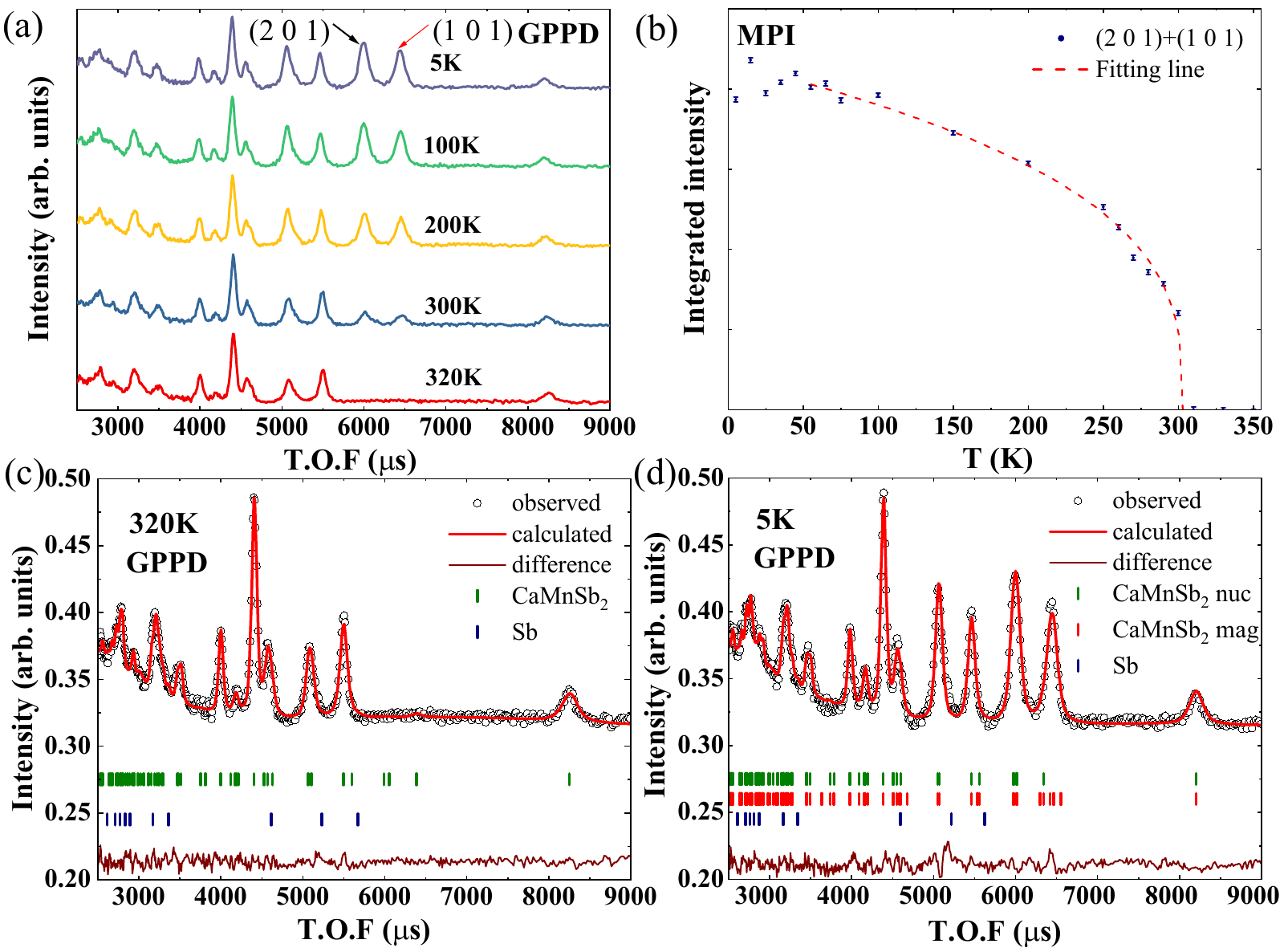}
	\caption{(a) Neutron diffraction patterns of CaMnSb$_2$ collected on GPPD at different temperatures, showing the emergences of (1 0 1) and (2 0 1) magnetic reflections upon cooling. (b) The temperature dependence of the magnetic diffraction intensity ($I\rm_M$), by integrating the total intensity of (1 0 1) and (2 0 1) reflections measured on MPI. The dashed line from 55 K to 303 K represents the fitting using a power law. (c)-(d) The Rietveld refinements to the neutron diffraction patterns collected on GPPD at 320 K and 5 K, respectively. The black open circles represent the observed intensities, and the calculated patterns according to the refinements are shown as red solid lines. The differences
between the observed and calculated intensities are plotted at the bottom as brown solid lines. The olive, red, and navy vertical bars indicate the nuclear reflections from CaMnSb$_2$, magnetic reflections from CaMnSb$_2$, and nuclear reflections from Sb impurity, respectively.
	}
\end{figure*}

\begin{figure}
	\centering
	\includegraphics[scale=0.78]{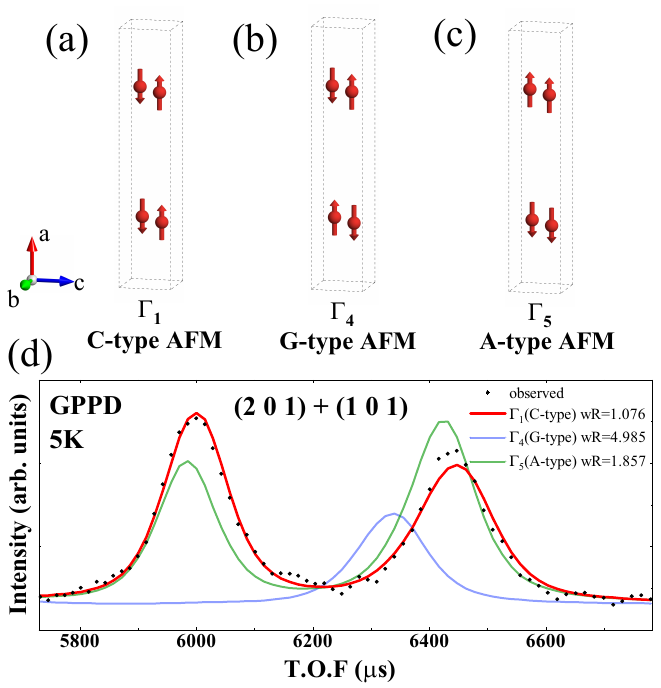}
	\caption{The spin configurations in different possible structures including the C-type AFM (a), G-type AFM (b) and A-type AFM (c), as described by the IR of $\Gamma_1$, $\Gamma_4$, and $\Gamma_5$, respectively, and the refinements to the (1 0 1) and (2 0 1) reflections using them accordingly (d). wR, the weighted magnetic R factor, is a metric used to evaluate the goodness of the fitting to the magnetic diffraction intensities, where a lower wR value implies a better agreement between the neutron diffraction pattern and the magnetic structure model.
	}
\end{figure}

Fig. 2(a) depicts the temperature dependence of the dc magnetic susceptibility ($\chi(T)$) of single-crystal CaMnSb$_2$, measured in the magnetic field of 1 T applied along the $a$ axis ($B$$\parallel$$a$) or the $bc$ plane ($B$$\bot$$a$), respectively. Similar to the behaviors reported in Ref. \onlinecite{23}, $\chi(T)$ exhibits a pronounced slope change at $T\rm_N$ = 303(1) K. Just below $T\rm_N$, $\chi(T)$ shows a clear decrease for $B$$\parallel$$a$ but stays almost invariant for $B$$\bot$$a$, suggesting the AFM nature of the magnetic order and the easy axis along the $a$ direction. In addition, a kink was observed in $\chi_{a}(T)$ around 270 K, in line with an earlier report, which was argued to arise from domain or disorder effect there \cite{23}. In addition, clear dHvA oscillations superimposed on a nonlinear background are observable in the isothermal magnetization ($M(B)$) curves measured at different temperatures for $B$$\parallel$$a$, as shown in Fig. 2(b), supporting the metallic nature of CaMnSb$_2$. The nonlinear background is likely to arise from minor ferromagnetic impurities introduced in the crystal growth.

After subtracting the nonlinear background $M\rm_{bac}$ in the $M(H)$ curves, the oscillatory parts $\Delta M$ = $M-M\rm_{bac}$ are plotted in Fig. 3(a) as a function of the reciprocal of the magnetic field (\(1/B\)). As the dHvA oscillations can be effectively described by the Lifshitz-Kosevich formula \cite{34}:
\[
{\Delta}M\propto{-B^{1/2}}{R\rm_T}{R\rm_D}\sin\left[2\pi\left(\frac{F}{B}-\gamma\right)\right],
\]
\[
{R\rm_T}=\frac{2\pi^2{k\rm_B}T{m_c}}{\hbar e B}\sinh\left(\frac{2\pi^2{k\rm_B}T{m_c}}{\hbar e B}\right),
\]
\[
{R\rm_D}=\exp\left(-\frac{2\pi^2{k\rm_B}{T\rm_D}{m_c}}{\hbar e B}\right),
\]
in which $R\rm_T$ and $R\rm_D$ are the thernal damping term and Dingle damping term, respectively, with $m_c$ being the effective cyclotron effective mass and $T\rm_D$ being the Dingle temperature. $\Delta M$ can be fitted to understand the properties of the Fermi surface. Utilizing the fast Fourier transform (FFT), two fundamental frequencies, \(F_\alpha = 9.2 \, \text{T}\) and \(F_\beta = 16.1 \, \text{T}\), were extracted (see Fig. 3(b)). By fitting the temperature dependence of the oscillating amplitudes (Amp.), as shown in Fig. 3(c), very small values of the cyclotron effective masses ($m_c$) of \(0.029 \, m_e\) and \(0.034 \, m_e\), respectively, can be estimated for the two frequencies.  
Furthermore, a nontrivial Berry phase $\phi\rm_B$ of 0.88 $\pi$ (close to $\pi$) is obtained according to $\gamma$ = 1/2$-$ $\frac{\phi\rm_B}{2\pi}$  \cite{35}. The very small values of $m_c$ and a nontrival $\phi\rm_B$ strongly support the scenario of nearly massless Dirac fermions hosted in CaMnSb$_2$. This is similar to the cases in SrMnBi$_2$, CaMnBi$_2$, BaMnBi$_2$, BaMnSb$_2$ and consistent with a previous study on CaMnSb$_2$ \cite{19, 23, 36, 37, 38}.

\begin{table*}	
	\caption{Nonzero IRs together with basis vectors ($\psi_v$) for Mn atoms in CaMnSb$_2$ with space group $Pnma$ and \textbf{$k=0$} propagation vector, obtained from irreducible representational analysis. The Mn atoms are defined as Mn No.1 (0.7499, 0.7500, 0.7280), Mn No.2 (0.7501, 0.2500, 0.2280), Mn No.3 (0.2501, 0.2500, 0.2720), Mn No.4 (0.2499, 0.7500, 0.7720), respectively.}
	\label{tab3}
	\begin{tabular}
		{
			p{1cm}
			>{\centering\arraybackslash}p{1cm}
			>{\centering\arraybackslash}p{3cm}
			>{\centering\arraybackslash}p{3cm}
			>{\centering\arraybackslash}p{3cm}
			>{\centering\arraybackslash}p{3cm}
			>{\centering\arraybackslash}p{3cm}
		}
		\toprule [1pt]
		IRs & $\psi_v$ & Component & Mn No.1 & Mn No.2 & Mn No.3 & Mn No.4 \\
		\midrule [1pt]
		$\Gamma_{1}$ & $\psi_{1}$ & Real & (1 0 0) & ($-$1 0 0) & ($-$1 0 0) & (1 0 0) \\[2pt]
		& $\psi_{2}$ & Real & (0 0 1) & (0 0 1) & (0 0 $-$1) & (0 0 $-$1) \\[2pt]
		$\Gamma_{2}$ & $\psi_{1}$ & Real & (0 1 0) & (0 $-$1 0) & (0 1 0) & (0 $-$1 0) \\[2pt]
		$\Gamma_{3}$ & $\psi_{1}$ & Real & (0 1 0) & (0 $-$1 0) & (0 $-$1 0) & (0 1 0) \\[2pt]
		$\Gamma_{4}$ & $\psi_{1}$ & Real & (1 0 0) & ($-$1 0 0) & (1 0 0) & ($-$1 0 0) \\[2pt]
		& $\psi_{2}$ & Real & (0 0 1) & (0 0 1) & (0 0 1) & (0 0 1) \\[2pt]
		$\Gamma_{5}$ & $\psi_{1}$ & Real & (1 0 0) & (1 0 0) & ($-$1 0 0) & ($-$1 0 0) \\[2pt]
		& $\psi_{2}$ & Real & (0 0 1) & (0 0 $-$1) & (0 0 $-$1) & (0 0 1) \\[2pt]
		$\Gamma_{6}$ & $\psi_{1}$ & Real & (0 1 0) & (0 1 0) & (0 1 0) & (0 1 0) \\[2pt]
		$\Gamma_{7}$ & $\psi_{1}$ & Real & (0 1 0) & (0 1 0) & (0 $-$1 0) & (0 $-$1 0) \\[2pt]
		$\Gamma_{8}$ & $\psi_{1}$ & Real & (1 0 0) & (1 0 0) & (1 0 0) & (1 0 0) \\[2pt]
		& $\psi_{2}$ & Real & (0 0 1) & (0 0 $-$1) & (0 0 1) & (0 0 $-$1) \\[2pt]
		\bottomrule [1pt]
	\end{tabular}
\end{table*}

\begin{table}[]
	\caption{Relative energies (in units of meV) of different magnetic configurations calculated for CaMnSb$_2$, without and with the effect of SOC being considered.}
	\label{tab4}
\centering
	\begin{tabular}
		{
			>{\small}p{2cm}
			>{\small}>{\centering\arraybackslash}p{2cm}
			>{\small}>{\centering\arraybackslash}p{2cm}
			>{\small}>{\centering\arraybackslash}p{2cm}
		}
		\toprule [1pt]
		Magnetic conf.		& C-type AFM & G-type AFM   & A-type AFM  \\
		\midrule [1pt]
		Without SOC & 0      & 0.572 & 173.7 \\
		With SOC    & 0      & 0.575 & 169.4 \\
		\bottomrule [1pt]
	\end{tabular}
\end{table}

Furthermore, neutron diffraction experiments on polycrystalline CaMnSb$_2$ were performed to deduce its magnetic structure. As shown in Fig. 4(a), the (1 0 1) and (2 0 1) reflections, which are forbidden nuclear peaks, emerge as magnetic peaks upon cooling, indicating a magnetic propagation vector of $k$ = 0. By integrating the total intensity of these two reflections, the temperature dependence of the magnetic diffraction intensity ($I\rm_M$) is plotted in Fig. 4(b). It is clear that $I\rm_M$ reaches the saturation below 55 K, suggesting a full ordering of the Mn moments. By fixing the N\'{e}el temperature as 303(1) K, the value determined by magnetization measurements, fitting to the $I\rm_{M}$$(T)$ behavior above 55 K using the powder law $I\rm_M$ $\propto(M)^2\propto(1-T/T\rm_N)^{2\beta}$ yields an exponent $\beta$ = 0.164(16). It is worthing pointing out that the magnetic order parameter seems to display a kink around 270 K, which coincides with the anomaly also observed around the same temperature in the magnetization data shown in Fig. 2(a). Definitely this anomaly can not be ascribed to the formation of domains, as argued in Ref. \onlinecite{23}, since neutron powder diffraction will average the domain effects. It might be due to a spin reorientation of the Mn moments occuring at an intermediate temperature, which needs to be clarified by further neutron measurements with better statistics. Our data collected around 270 K are insufficient to conclude such a possible spin canting.

\begin{figure*} [htb]
	\centering
	\label{DFT_band}
	\includegraphics[scale=1.5]{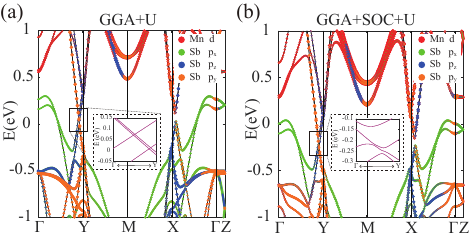}
	\caption{ Band structures of CaMnSb$_2$ calculated without (a) and with (b) the effect of SOC being considered. The insets in (a) and (b) show the linear dispersions around four crossing points and small gaps induced by SOC, respectively.}	
\end{figure*}

As shown in Fig. 4(c), the diffraction pattern at 320 K can be well refined by consideration of the CaMnSb$_2$ main phase (95.3\% wt) and a minor impurity phase of remaining Sb flux (4.7\% wt). As Sb is non-magnetic, its presence has no impact on our magnetic structure analysis. According to the irreducible representation analysis performed using the BASIREPS program integrated in the FULLPROF suite \cite{39},  for the case of $k$ = 0 and the space group of $Pnma$, the magnetic representations ($\Gamma\rm_{mag}$) for the Mn (4c) site can be decomposed as the sum of eight irreducible representations (IRs):
\[
\Gamma_{\mathrm{mag}}=2\Gamma_{1}^{1}\oplus\Gamma_{2}^{1}\oplus\Gamma_{3}^{1}\oplus2\Gamma_{4}^{1}\oplus2\Gamma_{5}^{1}\oplus\Gamma_{6}^{1}\oplus\Gamma_{7}^{1}\oplus2\Gamma_{8}^{1},
\]
whose basis vectors are listed in Table 3. Obviously, only the basis vector $\psi_{1}$ of three IRs correspond to AFM structures with an out-of-plane easy axis is consistent with the magnetization data shown in Fig. 2(a). As illustrated in Fig. 5(a), (b) and (c), $\Gamma_1$, $\Gamma_4$, and $\Gamma_5$ represent the C-type AFM, G-type AFM, and A-type AFM structures with the Mn moments aligned along the $a$ axis, respectively. Clearly, Fig. 5(d) shows a much better agreement between the calculated intensity adopting the model represented by $\Gamma_1$ and the observed intensities at 5 K in the low-$Q$ region where the magnetic form factor dominates, compared with those given by $\Gamma_4$ and $\Gamma_5$.  Thus, our neutron diffraction results unambiguously confirm the C-type AFM structure with the moment ferromagnetically coupled along the $a$ axis as the true magnetic structure of CaMnSb$_2$ (see Fig. 1(a)), similar to CaMnBi$_2$ and YbMnSb$_2$ \cite{12, 16}. The refined ordered moment of Mn is 3.167(13) $\mu\rm_{B}$ at 5 K, which is significantly smaller than the value of 5 $\mu\rm_{B}$ as expected for localized Mn$^{2+}$ ($3d^5, S=5/2$) moments but comparable to the values determined for other \textit{A}Mn\textit{X}$_2$ (\textit{A} = Ca, Sr, Ba, or Yb; \textit{X} = Sb or Bi) compounds  \cite{12, 13, 14, 15, 16, 17}, suggesting the itinerant nature of the 3$d$ electrons in such semimetals. 

The validity of the experimentally determined C-type AFM structure is further supported by our DFT calculations about the energies of different magnetic configurations, as shown in Table \ref{tab4}, in which the C-type AFM structure is clearly energetically more stable compared with the G-type and A-type cases, no matter whether or not the effect of spin-orbit coupling (SOC) is considered.

With the C-type AFM structure, the previous discrepancy regarding the topological properties of between CaMnSb$_2$ between Ref. \onlinecite{25} and Ref. \onlinecite{23, 24} can be settled now. C-type AFM structure maintains a $\mathcal{P}\times\mathcal{T}$ symmetry, while the G-type case assumed in the calculations of Ref. \onlinecite{25} only has the $\mathcal{P}$ symmetry. As the Dirac point requires 4-fold degeneracy, the G-type case cannot yield a nontrivial answer. On the other hand, the C-type AFM structure with the $\mathcal{P}\times\mathcal{T}$ symmetry makes it possible for CaMnSb$_2$ to be a nontrivial Dirac semimetal candidate, as suggested by Ref. \onlinecite{23, 24}.  Figs. 6(a) and 6(b) show our calculated band spectrum of CaMnSb$_2$ adopting the C-type AFM structure with and without the effect of SOC being considered, respectively. Four band crossings near the Y point in momentum space are observed. In the vicinity of these touching points, the dispersion of the Sb $5p_{y,z}$ bands exhibit linear behaviors, consistent with previous ARPES experiments \cite{24}. Due to the SOC effect, the touching points are slightly gapped, as shown in Fig. 6(b). Combined with the $\mathcal{P}\times\mathcal{T}$ symmetry, CaMnSb$_2$ is highly likely to be a Dirac semimetal.

We further propose that the quantum geometry and band property of CaMnSb$_{2}$ can be actually probed by a nonlinear anomalous Hall effect (NHE). With a $\mathcal{P}\times\mathcal{T}$
symmetry, Berry curvature in the momentum space is forced to vanish.
Nevertheless, an intrinsic second-order NHE has been predicted and
observed in $\mathcal{P}\times\mathcal{T}$-symmetric antiferromagnets \cite{40, 41, 42, 43, 44, 45, 46}, since it originates
from the Berry connection polarizability \cite{42}, which is related to the quantum metric dipole rather than the Berry curvature \cite{41, 47}.  
The intrinsic NHE not only provides a powerful tool for 
detecting band geometric quantities but also shows a promising capability of
monitoring the reorientation of the N\'{e}el vector in AFM spintronics \cite{42, 44, 48}.
According to the symmetry analysis for magnetic point group $m'm'm'$ associated with the C-type AFM structure,
we find the nonzero intrinsic NHE susceptibility tensor defined by
$\chi_{ijk}\equiv\frac{\partial^{2}j_{i}}{\partial E_{j}E_{k}}$ in
CaMnSb$_{2}$ are $\chi_{xyz}$, $\chi_{yxz}$ and $\chi_{zxy}$.
The experimental set-up for $\chi_{zxy}$ (similar approach for $\chi_{xyz}$
and $\chi_{yxz}$ with rotating axes) can be implemented by injecting
a low frequency ac current $\bm{I}^{\omega}$ along a direction in
the $xy$ plane, not parallel with $x$ or $y$ axis, and measuring
the induced Hall voltage $V_{z}^{2\omega}$. Notably, the band (anti)crossings
in topological Dirac semimetals could induce significant contribution
to the Berry connection polarizability and probably enhance the intrinsic
NHE susceptibility \cite{49}. Hence, CaMnSb$_{2}$ offers a promising platform
to study the impact of topological property on nonlinear electrical
transports. The NHE manifestation of the possible spin canting around
the susceptibility kink around 270 K is another interesting topic to explore in future. 

\section{Conclusion}

In summary, neutron diffraction measurements on polycrystalline CaMnSb$_2$ pulverized from high-quality single-crystal samples confirm a long-range C-type AFM ordering below the N\'{e}el temperature of $T\rm_N$ = 303(1) K, with the Mn moments aligned along the $a$ axis. Our DFT calculations support the validity of the C-type AFM structure, and reveal band crossings near the Y point and linear dispersions of the Sb $5p_{y,z}$ bands. Together with the characterizations from De Haas-van Alphen (dHvA) oscillations, we identify CaMnSb$_2$ as a topologically non-trivial magnetic Dirac semimetal. The magnetic space group $Pn'm'a'$ preserves a $\mathcal{P}\times\mathcal{T}$ symmetry, probably hosting an intrinsic second-order nonlinear Hall effect. Therefore, CaMnSb$_2$ represents a promising platform to study the impact of topological property on nonlinear electrical transports in antiferromagnets.

\begin{acknowledgments}

This work is financially supported by the National Natural Science Foundation of China (Grant No. 12074023, 12304053, and 12174018), the Large Scientific Facility Open Subject of Songshan Lake (Dongguan, Guangdong), and the Fundamental Research Funds for the Central Universities in China.

\end{acknowledgments}

\end{document}